\documentclass[%
 aip,
 amsmath,amssymb,
 preprint,%
]{revtex4-1}

\usepackage{siunitx}
\usepackage{xcolor}
\usepackage{graphicx}
\usepackage{dcolumn}
\usepackage{bm}

\usepackage[utf8]{inputenc}
\usepackage[T1]{fontenc}
\usepackage{mathptmx}

\begin{document}

\preprint{AIP/123-QED}

\title[Controlling skyrmion bubble confinement by dipolar interactions]{Controlling skyrmion bubble confinement by dipolar interactions\\}



\author{Fanny C. Ummelen}
\author{Tom Lichtenberg}%
    \email{t.lichtenberg@tue.nl}
\author{Henk J.M. Swagten}%
 \author{Bert Koopmans}
 \affiliation{ 
Department of Applied Physics, Eindhoven University of Technology, 5600 MB Eindhoven,
the Netherlands 
}%
\date{\today}

\begin{abstract}
Large skyrmion bubbles in confined geometries of various sizes and shapes are investigated, typically in the range of several micrometers. Two fundamentally different cases are studied to address the role of dipole-dipole interactions: (I) when there is no magnetic material present outside the small geometries and (II) when the geometries are embedded in films with a uniform magnetization. It is found that the preferential position of the skyrmion bubbles can be controlled by the geometrical shape, which turns out to be a stronger influence than local variations in material parameters. In addition, independent switching of the direction of the magnetization outside the small geometries can be used to further manipulate these preferential positions, in particular with respect to the edges. We show by numerical calculations that the observed interactions between the skyrmion bubbles and structure edge, including the overall positioning of the bubbles, can be explained by considering only dipole-dipole interactions.
\end{abstract}

\maketitle

Magnetic skyrmions are whirls in the magnetization in which neighbouring spins are rotated with respect to each other with a specific chirality. They are less hindered by pinning sites or defects than magnetic domain walls (DWs), and their size can be in the order of nanometers. These properties make them suitable for data storage. For the envisioned skyrmion racetrack memory \cite{skyrmion2,skyrmion1,Tomasello2014}, the skyrmions are required to be present in small geometrically confined structures, instead of infinite sheets of material. Therefore, the interaction between skyrmions and the edge of the magnetic structure is crucial. In fact, this interaction is necessary to prevent skyrmions from being expelled from the track, it can stabilize skyrmions in absence of an external magnetic field \cite{Boulle2016, klaui}, assist in their formation \cite{Du2015, Iwasaki2013}, and by reducing the width of the track it could be possible to reduce the size of the skyrmion and hence to achieve larger data storage densities \cite{review_skyrmions}. \\
\indent In the research field on skyrmions, usually a distinction is made between a `compact skyrmion' and a `skyrmion bubble'. These objects share many properties, but the latter has typically a much larger size and has a constant magnetization at its core \cite{review_skyrmions}. Numerical and experimental work on compact skyrmion confinement show that there is indeed a repulsive interaction between skyrmions and sample edges that is a result of tilting of the magnetic moments at the edge, which is caused by the Dzyaloshinskii-Moriya interaction (DMI) \cite{Rohart2013,Meynell2014}. For skyrmion bubbles, dipolar interactions are paramount in their stabilization, and because these stray fields will change near the sample edge, it is intuitively expected that the edges will influence the skyrmion bubbles via this mechanism. Though this has been realized before \cite{skyrmion_hall_effect}, to our knowledge it has never been studied in any detail. \\
\indent In this work addressing dipolar interactions in confined structures, we first study (I) skyrmion bubbles in isolated circular, square and triangular geometries and explore to what extend the position of these skyrmion bubbles can be controlled by the sample shape. It is observed that the skyrmion bubbles are repelled by the structure edges and arrange themselves in configurations reflecting the symmetry of the confining geometry. Next, we study (II) these small geometries containing skyrmion bubbles when embedded in uniformly magnetized films with a different magnetic anisotropy inside and outside the geometry. By independently switching the outside magnetization, it is observed that the skyrmion bubbles are either repelled by the structure edge, or may be localized throughout the entire geometry, fully in line with the intuitive role of dipolar interactions. Finally, we calculate theoretically how skyrmion bubble stability changes in the vicinity of an edge due to a change in dipolar interactions and calculate the preferred positions of skyrmion bubbles in some shapes considering dipolar interactions only. The observed behaviour matches well with these calculations, indicating that dipolar interactions are dominant in determining the positions of skyrmion bubbles with respect to edges and to each other.
\begin{figure}
    \centering
    \includegraphics{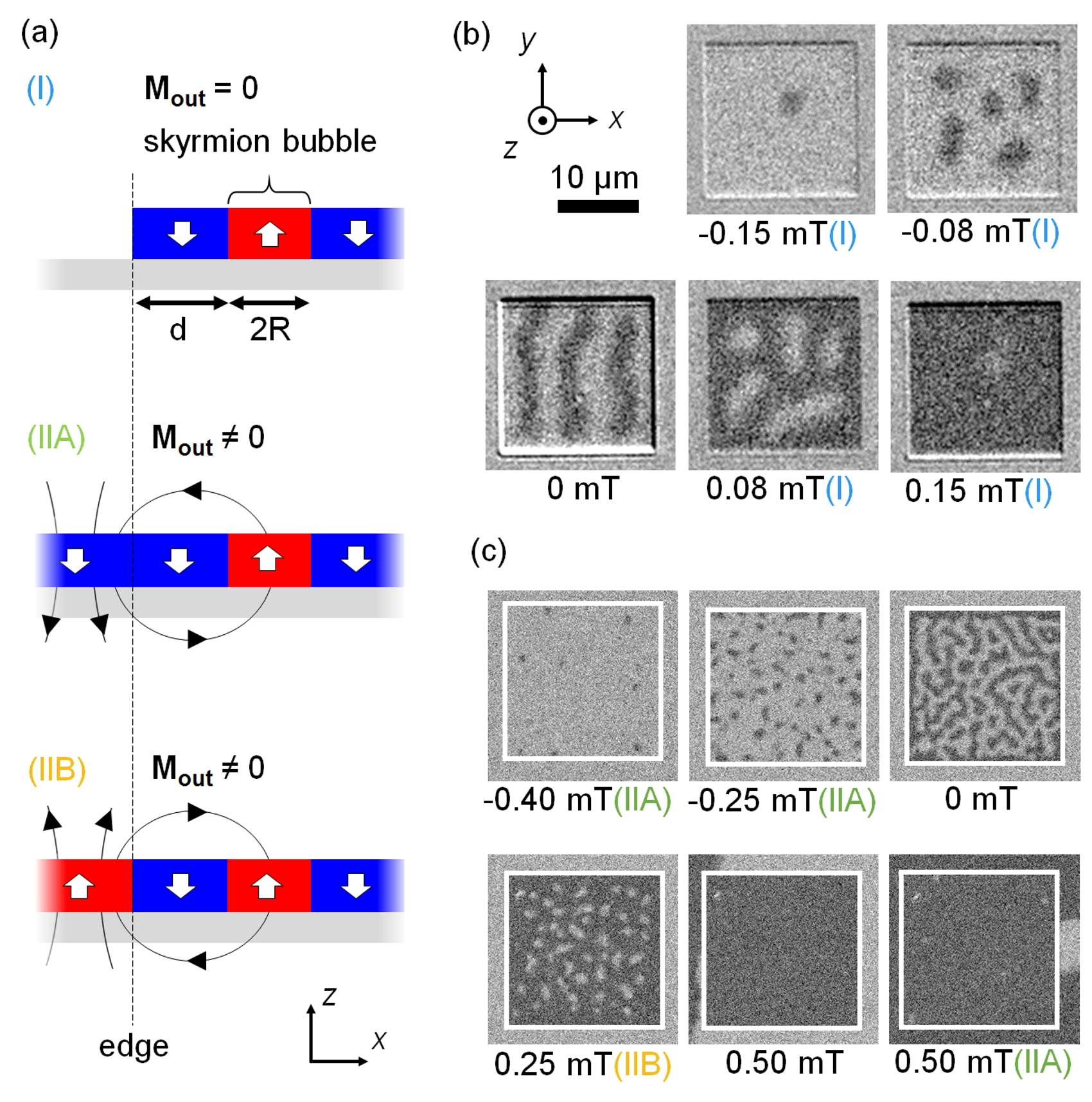}
    \caption{(a) Schematic representation of the three investigated edge types: I with no magnetic material outside the skyrmion containing geometry, II with the material outside the geometry magnetized (A) antiparallel to the skyrmion bubble core (B) parallel to the skyrmion bubble core. (b+c) Kerr microscope images of \SI{20}{\micro\metre} wide square structures for (b) a sample created by EBL (c) a sample created by FIB. Because the edge of the FIB structure is not visible, its location is indicated by a white frame. Below each image the applied perpendicular magnetic field and the corresponding edge type are indicated.}
    \label{C7:fig1}
\end{figure}
\\
\indent A Ta (\SI{5}{\nano\metre}) / Pt (\SI{4}{\nano\metre}) / Co ($t$) / Ir (\SI{4}{\nano\metre}) stack is used as a basis for the samples studied in this work. The two
different heavy metal layers adjacent to the magnetic layer
are known to induce large interfacial Dzyaloshinskii-Moriya
interaction\cite{skyrmionir}, which ensures that any DWs that are formed will
be of the Néel type and will have a fixed chirality. By careful tuning of the Co layer thickness ($t$), a balance between
the DW energy and dipolar energy can be found, such that
skyrmion bubbles can be stabilized using a small external
magnetic field\cite{schott,bubbles_Os,Ummelen2018}. Skyrmion bubbles are studied in circular, triangular, and square shapes of sizes ranging from \SI{4}{\micro\metre} to
to \SI{20}{\micro\metre}. The different structure sizes have different symmetries and thus enable us to investigate up to which dimensions
the edges influence the skyrmion bubbles, and down to which
dimensions skyrmion bubbles can be stabilized. Here we only
show the key results for a few of these structures that clearly
demonstrate the investigated bubble-edge interaction. Additional results are included in the supplementary material.\\
\indent Two different fabrication processes, one based on electron beam lithography (EBL) and one based on focused ion beam irradiation (FIB), are used (fabrication details are discussed in the supplementary material). With FIB, the anisotropy of a magnetic film can be controlled locally, allowing us to define regions in which skyrmion bubbles are stable. The two methods lead to two distinct situations at the edge of the structures. The EBL samples correspond to edge type (I), with no magnetic material outside the structure. The FIB samples correspond to edge type (II), with magnetic material outside of the investigated structures that has a homogeneous magnetization. In Fig.\ \ref{C7:fig1}(a) schematic side views of a skyrmion bubble near the edge is shown for these different edge types. \\
\indent Figure \ref{C7:fig1} also shows Kerr microscope images of a \SI{20}{\micro\metre} wide square created (b) by EBL and (c) FIB for various applied magnetic fields. The behavior as a function of magnetic field is comparable for both samples: at remanence a labyrinth domain structure forms, for small fields densely packed skyrmion bubbles occur, for increasingly larger fields only a few individual skyrmion bubbles remain until the uniform state is reached. The skyrmion bubbles in the EBL structure have different dimensions than in the FIB structure (the average radii are \SI{1.34}{\micro\metre} and \SI{0.7}{\micro\metre}, respectively) and the magnetic field at which these states occur is different for the two samples, suggesting a difference in the material parameters. Additionally, bubbles are observed at $t=0.7 \pm 0.1$ nm and $t=0.6 \pm 0.1$ nm for the EBL and FIB sample, respectively. Both samples show a property that is useful for our study: for the FIB sample it can be seen that at $\mu_0H_z = \SI{0.50}{\milli\tesla}$  the magnetization outside the irradiated structure switches. This coercive field is larger than the field for which the skyrmion bubbles are stabilized ($\mu_0H_z \approx \SI{0.25}{\milli\tesla}$). This makes it possible to study the behaviour of the skyrmion bubbles both when the magnetization outside the shape points parallel and antiparallel to the magnetization at the skyrmion core. For the EBL sample the dimensions of the bubbles and stripes are comparable to the size of the structure itself. The stripes at remanence are aligned with the edges of the structure \cite{Lee2008}, and for fields where skyrmion bubbles are stabilized, they are distributed such that the space in the structure is packed optimally. \\
\indent First, edge type (I) is explored by investigating the EBL sample. Observations on triangular shapes, which are highly anisotropic, of three different sizes are shown in Fig.\ \ref{C7:fig3}. Because the skyrmion bubbles exhibit thermal motion, spontaneous creation and annihilation, Kerr microscope movies are used 
to study their temporal variation instead of singular pictures. From these movies 100 consecutive frames are analyzed, extracting bubbles positions and sizes. The system is resaturated after every measurement. In Fig.\ \ref{C7:fig3} the positions of the skyrmion bubbles in the triangles for all 100 frames are indicated by semitransparent red dots, the size of which corresponds to the average size of a skyrmion bubble. The benefit of this representation is that if a bubble is detected at a certain spot multiple times, this spot becomes brighter red, and hence the preferential positions of the bubbles become visible.\\
\begin{figure}
\centering
\includegraphics{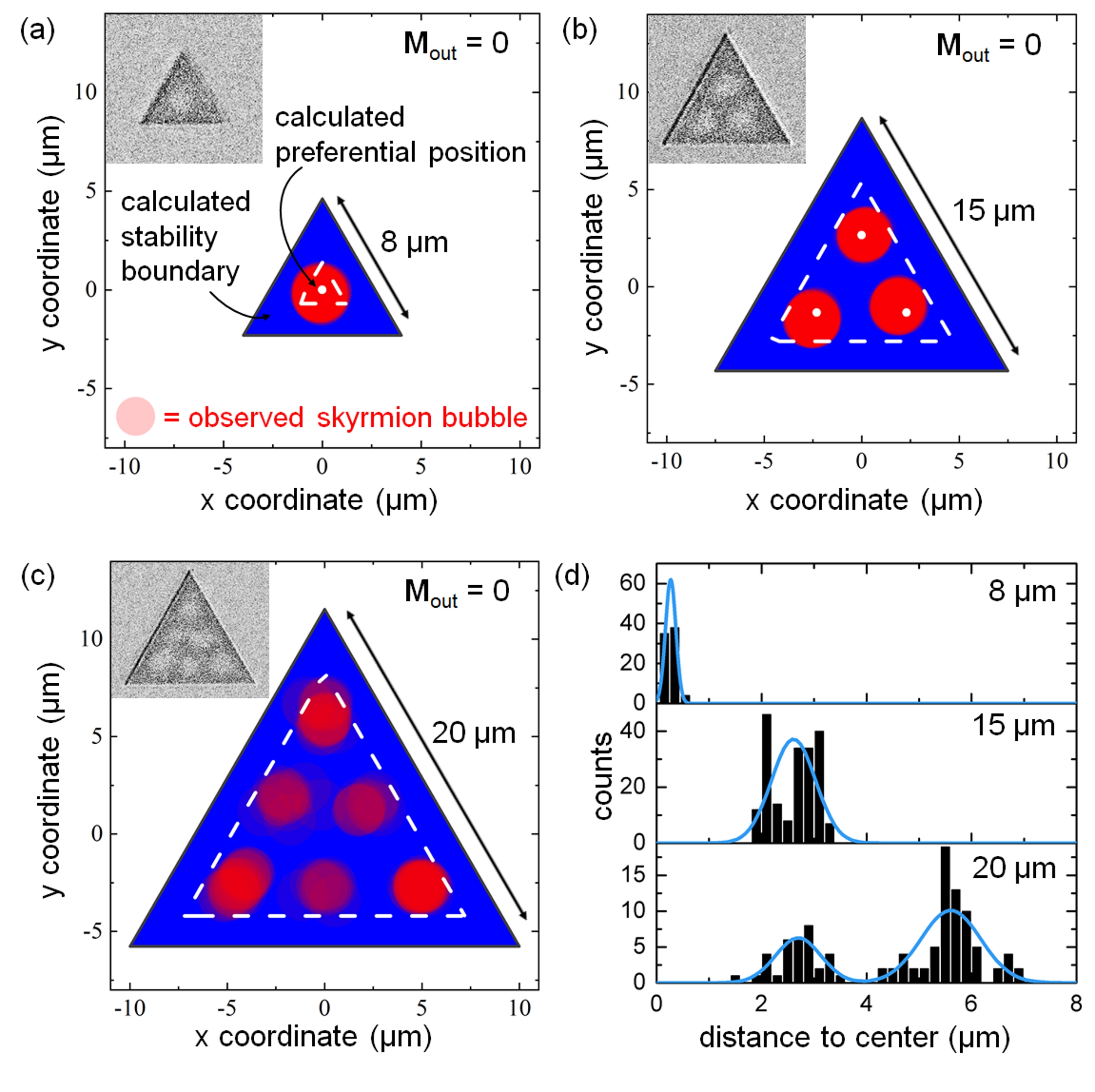}
\caption{Experimentally detected skyrmion bubbles in EBL-fabricated triangles with sides of (a) \SI{8}{\micro\metre} (b) \SI{15}{\micro\metre} (c) \SI{20}{\micro\metre} in 100 frames of a 15 frames per second video. The size of the red bubbles corresponds to the average bubble size. Insets show typical Kerr microscope images of the system. Dotted white lines indicate the stability region discussed later and white dots indicate the theoretically calculated bubble centers, which will be discussed later as well. (d) Histograms showing the number of detected skyrmion bubble as a function of the distance to the structure centre for all three triangles, together with Gaussian fits through the data.}
\label{C7:fig3}
\end{figure}
\indent Figure \ref{C7:fig3}(a) shows the results for the triangle with sides of \SI{8}{\micro\metre}, which is the smallest triangular structure in which we succeeded to stabilize skyrmion bubbles, at a field of \SI{0.05}{\milli\tesla}. Only one bubble is visible in each frame of the movie, and this bubble is always positioned at the centre of the triangle, as evidenced by the bright red spot. Fig. \ref{C7:fig3}(b) shows a triangle with sides of \SI{15}{\micro\metre} at a field of \SI{0.06}{\milli\tesla}. Three preferential positions are observed, which follow the triangular symmetry of the sample structure. However, the measured preferential positions are not completely symmetric, suggesting that local variations in material parameters also influence the preferential positions. In earlier works it was found that such local variations in material parameters were dominant in determining the equilibrium positions of skyrmions \cite{juge2018, zeissler2017}, but our results clearly indicate that the shape of the structures is the dominant influence for the magnetic structures investigated here. Last, Fig.\ \ref{C7:fig3}(c) shows the results for the triangle with \SI{20}{\micro\metre} wide sides at a field of \SI{0.05}{\milli\tesla}. Though in this case semitransparent spots dispersed throughout the structure are observed, which indicates that the skyrmion bubbles are now less strictly confined and move around more freely, six positions that are most preferred are clearly visible. In Fig.\ \ref{C7:fig3}(d) these visual observations are quantified as the the number of observed skyrmion bubbles as a function of the distance between the centre of the triangle and the centre of the bubble. For the \SI{8}{\micro\metre} structure, there are only counts in the close vicinity to the centre of the triangle. The small deviation from 0 can be explained either by an energy minimum due to local variation in material parameters or by the uncertainty in the detection of the structure edge during image analysis. For both the \SI{15}{\micro\metre} and \SI{20}{\micro\metre} triangle the peaks correspond to a triangular bubble distribution, showing that the bubbles are well confined within the structure.\\
\begin{figure}
\centering
\includegraphics{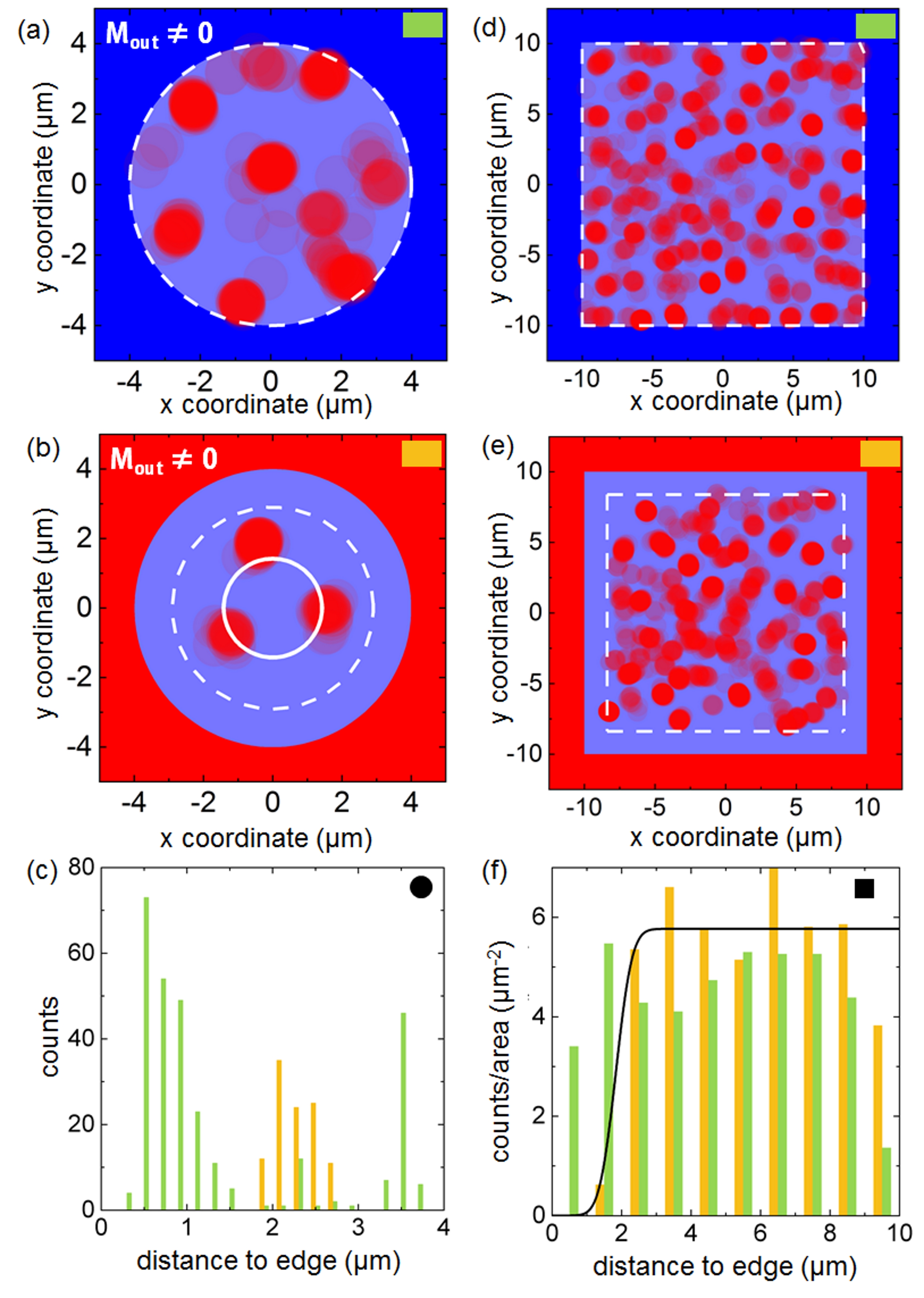}
\caption{Skyrmion bubble positions detected in the FIB circle with a \SI{8}{\micro\metre} diameter when the magnetization outside the structure points (a) antiparallel to (b) parallel to the magnetization at the core of the bubbles. We used 100 frames of a 10 frames per second Kerr microscope video. Light-blue areas indicate irradiated regions magnetized in the same direction as the blue areas. (d+e) Idem for the \SI{20}{\micro\metre} wide square. Dotted white lines indicate the stability region, the white solid line indicates the expected bubble positions. (c) and (f) show histograms of the number of detected bubbles for the situation shown in (a (green)+b (orange)) and (d (green)+e (orange)), respectively. The black line in (f) is a fit of the bubble density depicted in (e).}
\label{C7:fig2}
\end{figure}
\indent Next, the FIB sample (in particular the circle with a diameter of \SI{8}{\micro\metre}) is studied under influence of a \SI{0.25}{\milli\tesla} field, both for the situation that the magnetization outside of the FIB structure points antiparallel (edge type (IIA)) and parallel (edge type (IIB)) to the magnetization at the core of the skyrmion bubbles. Kerr microscope movies are analyzed in the same way as in the previous section and the results are plotted in Fig.\ \ref{C7:fig2}(a) and (b).
Because inside the structure containing the skyrmion bubbles the conditions are identical, it is remarkable that there is such a distinct difference between the preferential positions in (a) and (b). This difference is also apparent in Fig.\ \ref{C7:fig2}(c), which shows histograms with the number of observations as a function of the distance to the structure edge for both the situation in (a) (green) and (b) (orange). For situation (b) there are no observations closer than \SI{1.9}{\micro\metre} from the edge, which suggests a repelling force between the skyrmion bubbles and the structure edge. 
The fact that in Fig.\ \ref{C7:fig2}(a) there is a preferential spot in the middle of the structure that is not there in Fig.\ \ref{C7:fig2}(b) suggests that the interactions between the bubbles and the edge and the inter-bubble interactions are dominant over structural imperfections in determining the preferential spots. However, the data also suggests some influence of local variations in material properties, because if they were negligible the skyrmion observations would be distributed evenly along circles. \\
\indent For the \SI{20}{\micro\metre} sized squares, from which some raw images are shown in Fig.\ \ref{C7:fig1}(c), the observed skyrmion positions are shown in Fig.\ \ref{C7:fig2}(d) and (e), again for the situation that the magnetization outside the shape is aligned antiparallel or parallel to the cores of the skyrmions, respectively. The preferential positions seem to be distributed randomly through the FIB structure, indicating that the influence of the structure shape is no longer of relevance for this ratio between the structure size and skyrmion bubble size. However, in the vicinity of the edge the skyrmion bubbles can clearly be controlled by the magnetization outside the structure. Fig.\ \ref{C7:fig2}(f) quantifies the visual observations from Fig.\ \ref{C7:fig2}(d) (green) and (e) (orange): the number of bubbles counted per unit area is plotted as a function of the distance to the structure edge. For the green bars, the counts per area are indeed approximately constant as a function of the distance to the edge. For the orange bars this is not the case: a fit with an error function (black curve) reveals that the number of detected bubbles rapidly drops to zero around \SI{1.8}{\micro\metre} away from the edge.\\
\indent We will now discuss which mechanisms could be behind the observed interaction between skyrmion bubbles and the structure edge. Strong DMI has been reported for Pt/Co/Ir samples in literature, suggesting that edge states could play a role, just as for compact skyrmions. A problem with this interpretation for our observations is the length scale: the onset of this interaction is when the skyrmion and edge state `touch', so typically over the distance of the DW width and edge state width. These are in the order of tens of nanometers for the material stacks used here (supplementary material) while it is observed experimentally that the distance between the edge and skyrmion bubbles is in the order of micrometers. Therefore DMI-induced edge states cannot explain why our skyrmion bubbles are repelled by the structure edge.\\
\indent We use a combination of the thin wall model and numerical calculations to show that dipolar interactions are a plausible explanation for the observed results. The effect of a sample edge on a single bubble is studied by calculating the dipolar energy for different positions in a large magnetic film containing three edges visualized in fig. \ref{C7:fig1} (refer to the supplementary information for a more elaborate explanation). Fig.\ \ref{C7:fig1}(a) shows the dipolar fields that are involved for the three investigated edge types. Situation IIA shows the edge of a FIB structure where the magnetization beyond the edge is directed opposite to the magnetization at the bubble core. The stray fields emanating from beyond the edge help stabilize the bubble, and the dipolar energy should in principle be the same as for a bubble in an infinite film. Situation IIB shows a skyrmion bubble near the edge of a FIB structure, but now with the magnetization beyond the edge pointing in the opposite direction. The stray fields emanating from beyond the edge now increase the bubble energy. For edge type I, which corresponds to the EBL samples, there is no magnetic material and hence no stray field from beyond the edge. The bubble energy is now increased with respect to the energy of a bubble in an infinite film, because the dipolar fields that lower its energy are partially missing. Fig.\ \ref{C7:fig4} shows numerical calculations of how the dipolar energy varies as a function of the distance, $d$ (also indicated in Fig.\ \ref{C7:fig1}(a)) between the bubble and the edge for these three situations (see the supplementary information for details on this calculation).\\
\indent The stability and size of a skyrmion bubble can be calculated using the `thin wall model' \cite{bubble_materials, schott}. Here the energy of a circular domain in an infinite film is calculated with respect to the uniformly magnetized state. The size and stability of this circular domain is determined by the balance between the Zeeman energy, the DW energy and the dipolar energy. We determine the relevant material parameters experimentally, and within the margins of error a combination can be found resulting in stabilization of skyrmion bubbles with sizes as observed by Kerr microscopy (refer to the supplementary material). Near an edge, a bubble will feel a reduced dipolar energy, which can be included in the thin wall model to study bubble stability. For our material parameters, the dipolar term in the energy equation may be reduced about 6.6\% before the skyrmion bubble is no longer stable (see the inset in Fig.\ \ref{C7:fig4}(b), where the energy minimum disappears at this reduction). From the numerical calculations in the main figure, it can now be determined at what distance to the edge the dipolar energy is reduced by this amount. For the FIB edge this amounts to 1.0 bubble diameter and for the EBL edge 0.4 bubble diameters. The regions beyond which no skyrmion bubbles should be stable are indicated by dashed white lines in Fig.\ \ref{C7:fig3} and \ref{C7:fig2}. Indeed there are no observations beyond these limits, demonstrating that dipolar interactions are most likely responsible for the observed long range repulsion. Also, the same type of calculations can be used to predict the preferred skyrmion bubble positions in some of the geometries (again, see supplementary materials for details). These positions are indicated in white dots or lines in the data plots, and agree well with the experimentally observed bubble positions.
\begin{figure}
\centering
\includegraphics{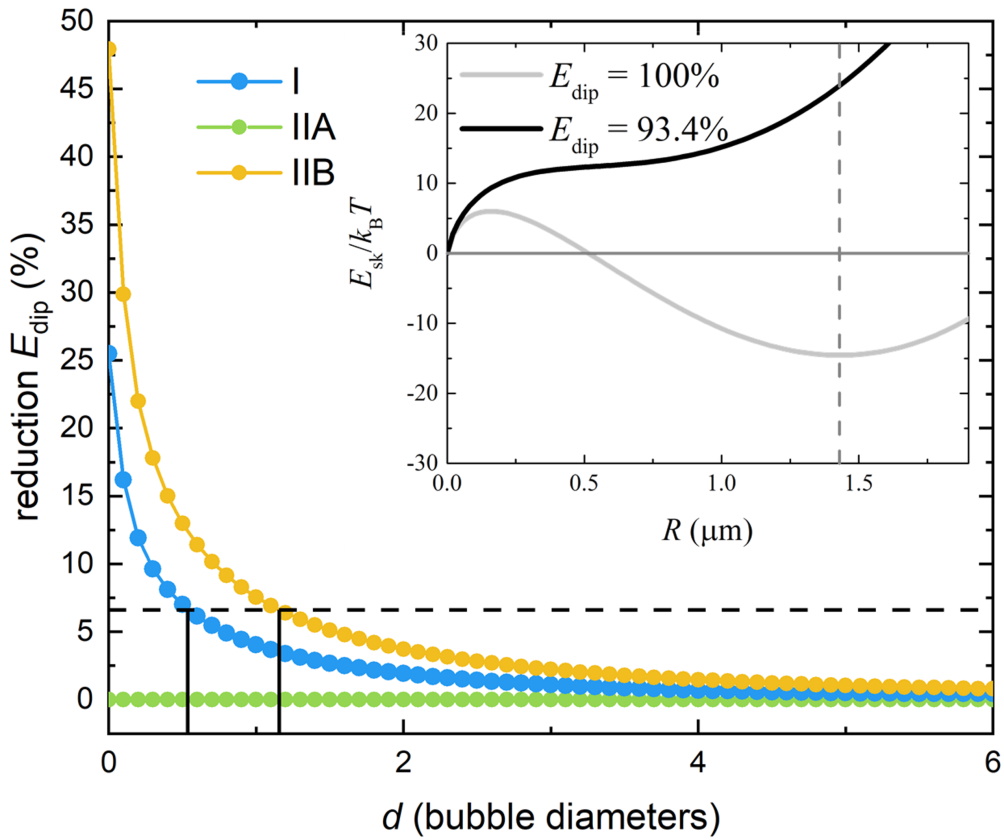}
\caption{Reduction in total bubble energy with respect to the situation of a single bubble in an infinite film  for the three investigated types of edges. Inset shows thin wall model calculations of the total skyrmion bubble energy, $E_{\textrm{tot}}$, as a function of its radius, $R$, both for standard and reduced dipolar energy contribution. Subsequently, the straight black lines indicate at what energy reduction and distance to the edge, $d$, skyrmion bubbles become unstable. For clarification, $R$ and $d$ are indicated in Fig.\ \ref{C7:fig1}.}
\label{C7:fig4}
\end{figure}
\\ \indent To summarize, we have investigated skyrmion bubbles in confined geometries created by two different techniques, which enables us to study how the bubbles are influenced by different types of structure edges. Dipolar interactions are the most plausible explanation for the observed repulsion between the bubbles and the edge. We have shown that ion beam irradiation can be used to confine bubbles in a novel way. The bubble-edge repulsion can be controlled by switching the magnetization outside the skyrmion-containing structures, posing exciting possibilities for future experiments and applications.\\
\indent See supplementary materials for (1) additional information on the material parameters, (2) a description of the used equipment and models, (3) an overview of the studied structures.\\
\indent This work is part of the research program of the Foundation for Fundamental Research on Matter (FOM), which is part of the Netherlands Organization for Scientific Research (NWO).
\bibliography{main}

\end{document}